\documentclass[twocolumn, aps, prc, superscriptaddress, showpacs, floatfix]{revtex4}
\usepackage[colorlinks=true,linkcolor=black,citecolor=blue, urlcolor=blue,bookmarks=false]{hyperref}
\hypersetup{breaklinks=true}
\usepackage[utf8]{inputenc}
\usepackage{graphicx}
\usepackage{natbib,slashed}
\usepackage{bm}
\usepackage{dcolumn}
\usepackage{array}
\usepackage{amsmath,mathtools}
\usepackage{amssymb}
\usepackage{multirow}
\usepackage{graphicx,subcaption}
\usepackage{tensor}
\usepackage{comment}
\usepackage{enumitem}
\graphicspath{{./fig/}}
\usepackage{float}
\usepackage[justification=raggedright,singlelinecheck=false]{caption}
\usepackage[english]{babel}
\usepackage[normalem]{ulem}  % \sout{old text} for strikeout
\usepackage{color} % For blue in-text comments and additions

\renewcommand{\sout}{\bgroup \color{red} \ULdepth=-.5ex \ULset}

\begin{document}
\title{Production of the $\phi$ and $\Omega$ via recombination of jet parton showers in relativistic heavy ion collisions}
\author{Kyong Chol Han}\email{kchan@pvamu.edu}
\affiliation{Department of Physics, Prairie View A$\&$M University, Prairie View, TX 77446, USA}
\affiliation{Institute for Quantum Science and Engineering, Texas A$\&$M University, College Station, TX 77843, USA}
\author{Su Houng Lee}\email{suhoung@yonsei.ac.kr}
\affiliation{Department of Physics and Institute of Physics and Applied Physics, Yonsei University, Seoul 03722, Korea}
\author{Sungtae Cho}\email[Corresponding author: ]{sungtae.cho@kangwon.ac.kr}
\affiliation{Division of Science Education, Kangwon National University, Chuncheon 24341, Korea}

\begin{abstract}
We study the production of the $\phi(1020)$ and $\Omega(1672)$ in
heavy ion collisions at $\sqrt{s_{NN}} = 5.02$ TeV by
employing two complementary approaches. In the first approach, the production of the $\phi$ and $\Omega$ is discussed in the coalescence model. In the second approach, we developed a hybrid
framework that combines the recombination of shower and thermal
partons with remnant string fragmentation.
Thermal partons in the quark-gluon plasma are modeled using a blast-wave
parameterization, while the phase space information of medium-modified parton showers is
generated from Q-PYTHIA, using unquenched jet partons obtained
from HIJING initial inputs.  We show that both approaches agree well
with the experimental measurements, and demonstrate that 
this hybrid framework provides deeper insight into the underlying strangeness components in the production of $\phi$ and $\Omega$ in relativistic heavy-ion collisions.
\end{abstract}

\maketitle

\section{Introduction}

Since the beginning of heavy ion collision experiments, hadrons
have been taken into account as useful probes to study the
properties of the quark-gluon plasma (QGP). In particular, strange
hadrons are considered important in examining the existence of the
QGP in early days in heavy ion collisions, revealing the
strangeness enhancement relative to elementary collisions
\cite{Rafelski:1982pu, Koch:1986ud}. The strangeness enhancement
has still been investigated through the production of
heavy-strange hadrons, such as charm-strange mesons
\cite{ALICE:2021kfc, Cho:2025lrc, Cho:2026xur}.

The study of hadron production in ultra-relativistic heavy-ion
collisions provides opportunities not only to obtain important insights
into the properties of the QGP, but also to understand the hadron
production mechanisms governing hadronization under extreme
conditions. 
The investigation of the enhanced transverse momentum
distribution ratio between the anti-proton and pion has led to
a better understanding of the hadron production mechanism by
coalescence from the QGP \cite{Greco:2003xt, Greco:2003mm,
Fries:2003vb, Fries:2003kq}.

Moreover, the hadron production mechanism in heavy ion collisions
has been more extensively studied, thereby enabling us to take into
account the hadronization of parton showers in jets using a hybrid
approach involving quark recombination and string fragmentation
\cite{Han:2016uhh}. In this hybrid model, perturbative parton showers
are turned into showers of constituent quarks and antiquarks via
gluon decay, and Monte Carlo methods are applied to recombine
quarks and antiquarks to form mesons or baryons based on
probabilities given by their overlap integrals with respect to the
meson or baryon Wigner functions, respectively. 
The remaining
constituent quarks are forced to be connected by a string through
the string fragmentation procedure in PYTHIA. This approach has
been successful in explaining the transverse momentum distribution
of pions, koans, and nucleons \cite{Han:2016uhh}.

In this hybrid model, it is also possible to distinguish where
each constituent quark comes from, i.e., from jet or shower
partons in heavy ion collisions, allowing us to identify the origin
of each component.  Thus, by adopting the hybrid model in studying
the production of the strange hadrons such as the $\phi(1020)$ and
$\Omega^-(1672)$, one can investigate the contribution of the
strange quarks in more detail in heavy ion collisions.
Nevertheless, the discussion in a hybrid approach so far has not
included the production of these strange hadrons, i.e., $\phi$
meson and $\Omega$ baryon. Furthermore, the experimental
measurement of $\Omega$ production in heavy ion
collisions at $\sqrt{s_{NN}}=5.02$ TeV has recently been made
\cite{ALICE:2025cqy}. 
Therefore, it is necessary to study the
production of the $\phi$ and $\Omega$ to understand not only the
hadron production mechanism but also to present theoretical
evaluations on the production of the $\Omega$ in heavy ion
collisions at $\sqrt{s_{NN}}=5.02$ TeV.

% Nevertheless, the discussion in a hybrid approach has not included
% the production of strange hadrons, i.e., $\phi$ meson and $\Omega$
% baryon. Considering that it is also possible to distinguish where
% each constituent quark comes from in the hybrid approach, one can
% identify the origin of each component in more detail, enabling to
% investigate the contributions originated from the strangeness
% enhancement in heavy ion collisions. finding i.e., whether when
% hadrons are formed is which constituent quarks when the hadron is
% produced

The $\phi$ meson is a particularly valuable probe of the QGP
because it is composed exclusively of strange valence quarks and
possess a relatively small hadronic interaction cross section and a  relatively long lifetime, with a width of about 4.25 MeV
\cite{ParticleDataGroup:2024cfk}, compared to the presumed lifetime of the hadronic stage in heavy ion collisions.
Consequently, the $\phi$ undergoes only limited rescattering
during the hadronic stage and largely preserve information from
the early partonic phase of the collision \cite{Rafelski:1982pu,
Shor:1984ui, ALICE:2021ptz}.

For similar reasons, extensive work is underway to
probe its properties at finite density produced in proton-nucleus
collisions through its $e^+e^-$ and $K^+K^-$ decay channels at the
J-PARC E16~\cite{Aoki:2024ood, PARCE16:2026pme} and
E88~\cite{Sako:2026ajt} experiments, respectively. Furthermore,
the recently observed directed flow of the $\phi$ meson in
heavy-ion collisions provides clues to its production mechanism
and properties at high baryon density~\cite{STAR:2026ley}.

The $\Omega$ baryon containing three strange quarks also serves as
a sensitive probe of strangeness production and equilibration in
the QGP \cite{Koch:1986ud}. In particular, it would be interesting to
investigate which strange quarks come from the thermal partons and
which strange quarks come from the shower partons when the $\Omega$ is formed in heavy ion collisions. Moreover, the $\phi$
and $\Omega$ are mostly produced from strange quarks in the QGP as
they only have their feed-down contributions from heavier
charm-strange hadrons, such as the $D_s$ and $\Omega_c$
\cite{ParticleDataGroup:2024cfk}. 
Thus, one can examine more clearly the origin of constituent quarks with a hybrid model for the production of the $\phi$ and $\Omega$, and, by comparing such theoretical calculations with measurements of the production of the $\phi$ and $\Omega$ at the Large Hadron Collider (LHC), one would be able to provide stringent constraints on theoretical models of hadronization and the evolution of the deconfined medium.

For this reason, we investigate in this work the production of the
$\phi$ meson and the $\Omega$ baryon in central Pb-Pb collisions
at $\sqrt{s_{NN}}=5.02$ TeV at the LHC. To explore the underlying
hadronization dynamics, we consider two complementary approaches. 
Starting from the discussion on their production in the coalescence model, we investigate the transverse momentum distributions of $\phi$ and $\Omega$ produced via recombination of jet parton shower in heavy ion collisions.

We extend the hybrid hadronization framework originally
developed in Ref.~\cite{Han:2016uhh} for the hadronization of
perturbative parton showers in vacuum, with applications to
pions, kaons, protons, and $\Lambda$ baryons, to the production
of $\phi$ and $\Omega$ in a QGP medium. In this extension,
thermal partons from the QGP are combined with
medium-modified shower partons originating from energetic jets
through quark recombination, while shower partons that remain
unrecombined are subsequently hadronized through the Lund string
fragmentation model. This event-by-event hybrid framework
provides a unified description of hadron production over a
broad transverse-momentum range by incorporating both
recombination and fragmentation.

We calculate the transverse-momentum spectra of the $\phi$ meson
and $\Omega$ baryon, as well as the $\Omega/\phi$ yield ratio, and compare the calculated spectra with measurements from the
ALICE Collaboration. These results allow us to examine the
relative contributions of thermal and shower partons to
strange-hadron production and to investigate the transition
between recombination and fragmentation in the hadronization of
deconfined quarks in relativistic heavy-ion collisions. All
calculations presented in this work are performed on an
event-by-event basis.

This paper is organized as follows. In Sec.~\ref{sec2}, we discuss
the production of the $\phi$ and $\Omega$ in the coalescence
model, and presents their transverse momentum distributions and
yields. Sec. ~\ref{sec3} is devoted to the description of the
$\phi$ and $\Omega$ production in a hybrid model, starting from
explaining the generation of the spatial and momentum
distributions of thermal and shower partons produced in central
(0--10\%) Pb--Pb collisions at $\sqrt{s_{NN}}=5.02$ TeV to
introducing a Monte Carlo hybrid hadronization framework that
combines quark recombination with remnant string fragmentation.
Numerical results for the transverse-momentum spectra and the
$\Omega/\phi$ yield ratio are also presented and compared with
available experimental data in Sec. ~\ref{sec3}. Finally, we
summarizes the main findings and presents the conclusions in
Sec.~\ref{sec4}.

\section{Production of the $\phi$ and $\Omega$ by coalescence
in Relativistic Heavy-Ion Collisions}
\label{sec2}

We first investigate the production of the $\phi$ meson and
$\Omega$ baryon via the coalescence of strange quarks produced in
relativistic heavy-ion collisions. We consider the transverse
momentum distribution and yield of the $\phi$ and $\Omega$ based
on the coalescence model \cite{Greco:2003xt, Greco:2003mm}.

The yield of the meson produced from a quark and anti-quark,
denoted respectively by the subscript $q$ and $\bar{q}$ is given
by,
\begin{eqnarray}
&& N_M=g_M\int p_q\cdot d\sigma_q p_{\bar{q}} \cdot
d\sigma_{\bar{q}} \frac{d^3 p_q}{(2\pi)^3 E_q}\frac{d^3
p_{\bar{q}}}{(2\pi)^3 E_{\bar{q}}}
\nonumber \\
&& \qquad\times  f_q(r_q, p_q) f_{\bar{q}}(r_{\bar{q}},
p_{\bar{q}}) W_M(r_q, r_{\bar{q}}; p_q, p_{\bar{q}}).
\label{MesonCoalGeneral}
\end{eqnarray}
Similarly, the yield of the baryon produced from three quarks
denoted by the subscript $q_i$ with $i=1,2,3$ is given by,
\begin{eqnarray}
&& N_B=g_B\int p_{q_1}\cdot d\sigma_{q_1} p_{q_2}\cdot
d\sigma_{q_2} p_{q_3}\cdot d\sigma_{q_3} \nonumber \\
&& \qquad\times \frac{d^3 p_{q_1}}{(2\pi)^3 E_{q_1}} \frac{d^3
p_{q_2}}{(2\pi)^3 E_{q_2}} \frac{d^3
p_{q_3}}{(2\pi)^3 E_{q_3}} \nonumber \\
&& \qquad\times  f_{q_1}(r_{q_1}, p_{q_1})f_{q_2}(r_{q_2},
p_{q_2}) f_{q_3}(r_{q_3}, p_{q_3}) \nonumber \\
&& \qquad\times W_B(r_{q_1}, r_{q_2}, r_{q_3}; p_{q_1}, p_{q_2},
p_{q_3}), \label{BaryonCoalGeneral}
\end{eqnarray}
where $d\sigma_q$ is the space-like hypersurface element for a
quark, and $f_q(r_q, p_q)$ is a quark covariant distribution
function representing the number of those quarks $q$, $N_q$ in the
system satisfying $\int p_q\cdot d\sigma_q d^3
p_q/((2\pi)^3E)f_q(r_q, p_q)=N_q$. The subscript $M$ and $B$ in
Eqs. (\ref{MesonCoalGeneral}) and (\ref{BaryonCoalGeneral}) are
denoted by the meson and baryon, respectively. The factor $g_M$
and $g_B$ are the ratios of the degeneracy factors  of the the meson and baryon  to those of their constituent quarks,  respectively, e.g.,
$g_\phi=3/(2\cdot 3)^2$ and $g_\Omega=1/(2\cdot 3)^3$.

In the non-relativistic limit, Eqs. (\ref{MesonCoalGeneral}) and
(\ref{BaryonCoalGeneral}) in mid-rapidities are reduced to
\cite{Greco:2003mm, Greco:2003xt, Oh:2009zj}
\begin{eqnarray}
&& \frac{dN_{\phi}}{d\vec p_T}=\frac{g_{\phi}}{V} \int d\vec r
d\vec p_{sT} d\vec p_{\bar{s}T} \delta^{(2)}(\vec p_T-\vec
p_{sT}-\vec p_{\bar{s}T}) \nonumber \\
&& \qquad\quad\times\frac{dN_s}{d \vec p_{sT}} \frac{dN_{\bar{s}}}
{d\vec p_{\bar{s}T}} W_{\phi}(\vec r, \vec k),
\label{CoalTransphi}
\end{eqnarray}
for the $\phi$ meson and,
\begin{eqnarray}
&& \frac{dN_{\Omega}}{d\vec p_T}=\frac{g_{\Omega}}{V^2} \int d\vec
r_1 d\vec r_2 d\vec p_{s_1 T}d\vec p_{s_2 T}
d\vec p_{s_3 T} \nonumber \\
&& \qquad\quad \times \delta^{(2)}(\vec p_T-\vec p_{s_1 T}-\vec
p_{s_2 T}-\vec p_{s_3 T}) \nonumber \\
&& \qquad\quad \times\frac{dN_{s_1}}{d \vec p_{s_1 T}}
\frac{dN_{s_2}} {d\vec p_{s_2 T}} \frac{dN_{s_3}}{d\vec p_{s_3 T
}}W_{\Omega}(\vec r_1, \vec r_2, \vec k_1, \vec k_2),
\label{CoalTransOmega}
\end{eqnarray}
for the $\Omega$ baryon. In Eqs. (\ref{CoalTransphi}) and
(\ref{CoalTransOmega}) the assumption on the Bjorken correlation
between spatial and momentum rapidities, $\eta$ and $y$
respectively, or the boost-invariant longitudinal momentum
distributions for quarks satisfying $\eta=y$ has been made. $\vec
r$, $\vec r_1$, $\vec r_2$ and $\vec k$, $\vec k_1$, $\vec k_2$ in
Eqs (\ref{CoalTransphi}) and (\ref{CoalTransOmega}) are relative
distances and transverse momenta between strange quarks,
respectively, given by,

\begin{eqnarray}
&& \vec R_M=\frac{m_s\vec r_s+m_{\bar{s}}\vec
r_{\bar{s}}}{m_s+m_{\bar{s}}}, \qquad
\vec r=\vec r_s-\vec r_{\bar{s}}, \nonumber \\
&& \vec K_M=\vec p_{sT}'+\vec p_{\bar{s}T}', \qquad \vec
k=\frac{m_s\vec p_{\bar{s}T}'-m_{\bar{s}}\vec
p_{sT}'}{m_s+m_{\bar{s}}}, \label{relcoordmeson}
\end{eqnarray}
for the $\phi$ with the reduced mass, $\mu=m_s m_{\bar
s}/(m_s+m_{\bar s})$. Similarly,

\begin{eqnarray}
&& \vec R_B=\frac{m_{s_1}\vec r_{s_l}+m_{s_2}\vec r_{s_2}+m_{s_3}
\vec r_{s_3}}{m_{s_1}+m_{s_2}+m_{s_3}}, \nonumber \\
&& \vec r_1=\vec r_{s_1}-\vec r_{s_2}, \nonumber \\
&& \vec r_2=\frac{m_{s_1}\vec r_{s_1}+m_{s_2}\vec
r_{s_2}}{m_{s_1}+m_{s_2}}-\vec r_{s_3}, \nonumber \\
&& \vec K_B=\vec p_{s_1T}'+\vec p_{s_2 T}'+\vec p_{s_3 T}', \nonumber \\
&& \vec k_1=\frac{m_{s_2}\vec p_{s_1 T}'-m_{s_1}\vec p_{s_2
T}'}{m_{s_1}+m_{s_2}},
\nonumber \\
&& \vec k_2=\frac{m_{s_3}(\vec p_{s_1 T}'+\vec p_{s_2
T}')-(m_{s_1}+m_{s_2}) \vec p_{s_3T}'}{m_{s_1}+m_{s_2}+m_{s_3}},
\label{relcoordbaryon}
\end{eqnarray}
with corresponding reduced masses,

\begin{equation}
\mu_1=\frac{m_{s_1}m_{s_2}}{m_{s_1}+m_{s_2}}, \qquad
\mu_2=\frac{(m_{s_1}+m_{s_2})m_{s_3}}{m_{s_1}+m_{s_2}+m_{s_3}},
\label{reducedmass_Omega}
\end{equation}
for the $\Omega$.

The relative momentum $\vec k$ carries the information between two
frames, the fireball frame and the rest frame of the $\phi$ meson
via the Lorentz transformation from $\vec p_{\bar{s}T}$ and $\vec
p_{sT}$, the transverse momenta of quarks in the fireball frame to
$\vec p_{\bar{s}T}'$ and $\vec p_{sT}'$, those of quarks in the
rest frame of the $\phi$ meson, \cite{Scheibl:1998tk, Oh:2009zj}.
By the same token, the relative momenta $\vec k_1$ and $\vec k_2$
in Eq. (\ref{reducedmass_Omega}) connects the information between
the fireball frame and the rest frame of the $\Omega$ baryon
through Lorentz transformation of transverse momenta in the
fireball frame, $\vec p_{s_1T}$, $\vec p_{s_2T}$, and $\vec
p_{s_3T}$ to those in the rest frame of the $\Omega$, $\vec
p_{s_1T}'$, $\vec p_{s_2T}'$, and $\vec p_{s_3T}'$.

It should be noted that the yield and transverse momentum
distribution are shown to be independent of different choices for
relative coordinates and momenta between quarks inside the hadron,
if all quarks are in the ground state, suitable for the $s$-wave
Gaussian Wigner function with one oscillator frequencis $\omega$
\cite{Cho:2019syk}. Moreover, since the $\Omega$ baryon is made up
of the three same strange quarks, other choices for relative
configurations between three strange quarks rather than those
shown in Eq. (\ref{relcoordbaryon}), do not affect the yield or
transverse momentum distribution.

Adopting the $s$-wave Wigner function built from harmonic
oscillator wave functions, we have
\begin{eqnarray}
&& W_{\phi}(\vec r, \vec k)
=8e^{-\frac{r^2}{\sigma^2}-\sigma^2k^2} \nonumber \\
&& W_{\Omega}(\vec r_1, \vec r_2, \vec k_1, \vec k_2) \nonumber \\
&& \quad =8^2\exp{\bigg(-\frac{r_1^2}{\sigma_1^2}-\sigma_1^2k_1^2
\bigg)}\exp{\bigg(-\frac{r_2^2}{\sigma_2^2}-\sigma_2^2k_2^2\bigg)},\nonumber\\
\label{WignerS}
\end{eqnarray}
with the relative coordinates and transverse momenta for the
$\phi$, Eq. (\ref{relcoordmeson}) and those for the $\Omega$, Eq.
(\ref{relcoordbaryon}), respectively.

In Eq. (\ref{WignerS}), $\sigma$ is related to the oscillator
frequency in the harmonic oscillator wave function for the $\phi$,
$\omega_{\phi}$ with $\sigma^2=1/(\mu\omega_{\phi})$. Also
$\sigma_1$ and $\sigma_2$ are connected to the oscillator
frequency for the $\Omega$ wave function, $\omega_{\Omega}$ with
$\sigma_1^2=1/(\mu_1\omega_{\Omega})$ and
$\sigma_2^2=1/(\mu_2\omega_{\Omega})$, respectively.

After inserting the Wigner functions, Eq. (\ref{WignerS}) into
Eqs. (\ref{CoalTransphi}) and (\ref{CoalTransOmega}), and carrying
out the coordinate space integration, we obtain,

\begin{eqnarray}
&& \frac{dN_{\phi}}{d\vec p_T}=\frac{g_{\phi}}{V}
(2\sqrt{\pi}\sigma)^3 \int d\vec p_s d\vec p_{\bar{s}}
\delta^{(2)}(\vec p_T-\vec p_{sT}-
\vec p_{\bar{s}T}) \nonumber \\
&& \qquad\quad\times\frac{dN_s}{d \vec p_{sT}} \frac{dN_{\bar{s}}}
{d\vec p_{\bar{s}}} e^{-\sigma^2 k^2}, \label{CoalTransphiW}
\end{eqnarray}
for the $\phi$ meson and,
\begin{eqnarray}
&& \frac{dN_{\Omega}}{d\vec p_T}=\frac{g_{\Omega}}{V^2}
(2\sqrt{\pi})^6 (\sigma_1\sigma_2)^3 \int d\vec p_{s_1}d\vec
p_{s_2} d\vec p_{s_3} \nonumber \\
&& \qquad\quad \times \delta^{(2)}(\vec p_T-\vec p_{s_1 T}-\vec
p_{s_2 T}-\vec p_{s_3 T}) \nonumber \\
&& \qquad\quad \times\frac{dN_{s_1}}{d \vec p_{s_1}}
\frac{dN_{s_2}} {d\vec p_{s_2}} \frac{dN_{s_3}}{d\vec
p_{s_3}}\exp{\bigg(-\sigma_1^2 k_1^2-\sigma_2^2k_2^2\bigg)},
\nonumber \\
\label{CoalTransOmegaW}
\end{eqnarray}
for the $\Omega$ baryons.

In regard to the oscillator frequencies for the $\phi$,
$\omega_{\phi}$, we take the $\phi$ meson size of 0.46 fm for the
root-mean-square radius of the $\phi$ meson, $\sqrt{\langle
r^2\rangle_\phi}$ \cite{Chen:2006vc, Wang:2023wlq}, leading to the
oscillator frequency of the $\phi$ meson, $\omega_{\phi}=1.104$
GeV via the relation between the root-mean-square radius and
$\sigma$, $\langle r^2\rangle_\phi=3/2
\sigma^2=3/(2\mu\omega_\phi)$.

By the same token, we consider the root-mean-square radius of the
$\Omega$ for the oscillator frequency in the Wigner function on
the $\Omega$, $\langle r^2\rangle_\Omega$. The mean distance
between the center of mass among three strange quarks inside the
$\Omega$, Eq. (\ref{relcoordbaryon}) and the position of one of
strange quarks, e.g., the first strange quark, $\vec r_{s_1}$
becomes,

\begin{eqnarray}
&& \langle r^2\rangle_\Omega=\langle (\vec R-\vec r_{s_1})^2
\rangle \nonumber \\
&& \qquad=\frac{3}{2}\frac{1}{\omega_\Omega}\frac{m_{s_2}+m_{s_3}}
{(m_{s_1}+m_{s_2}+m_{s_3})m_{s_1}}. \label{rootmeansquareradius}
\end{eqnarray}
On the other hand, its mean charged radius becomes
\cite{Oh:2009zj},

\begin{eqnarray}
&& \langle r^2\rangle_{\Omega, ch} \nonumber \\
&&=\langle (\vec R-\vec r_{s_1})^2Q_{s_1}+(\vec R-\vec
r_{s_2})^2Q_{s_2}+(\vec R-\vec r_{s_3})^2Q_{s_3} \rangle \nonumber \\
&&=\frac{3}{2}\frac{1}{\omega_\Omega}\frac{1}{m_{s_1}+m_{s_2}+m_{s_3}}
\nonumber \\
&&\times\bigg(\frac{m_{s_2}+m_{s_3}}{m_{s_1}}Q_{s_1}+\frac{m_{s_3}
+m_{s_1}}{m_{s_2}}Q_{s_2}+\frac{m_{s_1}+m_{s_2}}{m_{s_3}}Q_{s_3}\bigg).
\nonumber \\
\label{menachargeradius}
\end{eqnarray}

As the $\Omega$ baryon consists of three strange quarks of the
same mass, $\langle (\vec R-\vec r_{s_1})^2 \rangle=\langle (\vec
R-\vec r_{s_2})^2 \rangle=\langle (\vec R-\vec r_{s_3})^2
\rangle$, and moreover, its mean charged radius, Eq.
(\ref{menachargeradius}) and root-mean-square radius, Eq.
(\ref{rootmeansquareradius}) becomes identical except the sign
coming from the charge of the strange quark.

We take the root-mean-square radius of the $\Omega$, $\langle
r^2\rangle_\Omega=0.18$ fm$^2$ evaluated in the non-relativistic
quark model \cite{Povh:1990ad}, leading to the oscillator
frequency of the $\Omega$, $\omega_\Omega=0.433$ GeV. This radius
is bigger than the mean charged radius obtained in the bound state
model, $\langle r^2\rangle_{\Omega,ch}=-0.16$ fm$^2$
\cite{Gobbi:1992mp} while it is smaller than that calculated in
the relativistic quark model, $\langle
r^2\rangle_{\Omega,ch}=-0.22$ fm$^2$ \cite{Ramalho:2009gk}. The
above harmonic oscillator frequency of the $\Omega$,
$\omega_\Omega=0.433$ corresponds to the $\sigma_{B1}$ and
$\sigma_{B2}$ \cite{Han:2016uhh}, 3.04 fm and 2.63 fm,
respectively.

For the transverse momentum distribution of strange quarks, we
adopt that obtained to explain the experimental measurement of the
$\phi$ meson at $\sqrt{s_{NN}}=5.02$ TeV at LHC
\cite{Cho:2025lrc},

\begin{eqnarray} && \frac{dN_s}{d \vec p_{sT}}=\left\{
\begin{array}{l}
\frac{V}{(2\pi)^3}m_Te^{-m_T/T_{\rm eff}}, \quad~~~p_{sT} \le 1.50~\textrm{GeV} \\
21.95 ~\mathrm{(GeV^{-2})} e^{-0.17(p_{sT}/p_{0T})^{3.23}} \\
+\frac{80112 ~\mathrm{(GeV^{-2})}}
{(1.0+(p_{sT}/p_{0T})^{0.65})^{10.29}}, ~p_{sT} >1.50~\textrm{GeV}
\end{array} \right. \nonumber \\
\label{d2NsdpT2}
\end{eqnarray}
where $g_s$, $V$, $m_T=\sqrt{p_{T}^2+m_s^2}$ and
$T_{\rm eff}$ are the degeneracy factor of a strange quark for its 
spin and color, $g_s$=6, the coalescence volume, the transverse mass with the mass of a strange quark, $m_s$ and the effective temperature,
respectively, with $p_{0T}$ being 1.0 GeV. We adopt here $m_s$=500 MeV and $T_{\rm eff}$=173 MeV \cite{Cho:2025lrc}. Due to the negligible baryon chemical potential at $\sqrt{s_{NN}}=5.02$ TeV energy at
LHC, it is assumed that the transverse momentum distribution of
anti-strange quarks is the same as that of strange quarks. Taking the coalescence volume 3360 fm$^3$ at $\sqrt{s_{NN}}=5.02$
TeV at LHC obtained from the entropy conservation condition during
the expansion of the system between the critical and hadronization
temperatures \cite{Cho:2025lrc}, we obtain from the above transverse momentum distribution of strange quarks in Eq. 
(\ref{d2NsdpT2}) about 780 for the total number of strange quarks at mid-rapidity in central collisions at $\sqrt{s_{NN}}=5.02$ TeV.

\begin{figure}[!t]
\begin{center}
\includegraphics[width=0.50\textwidth]{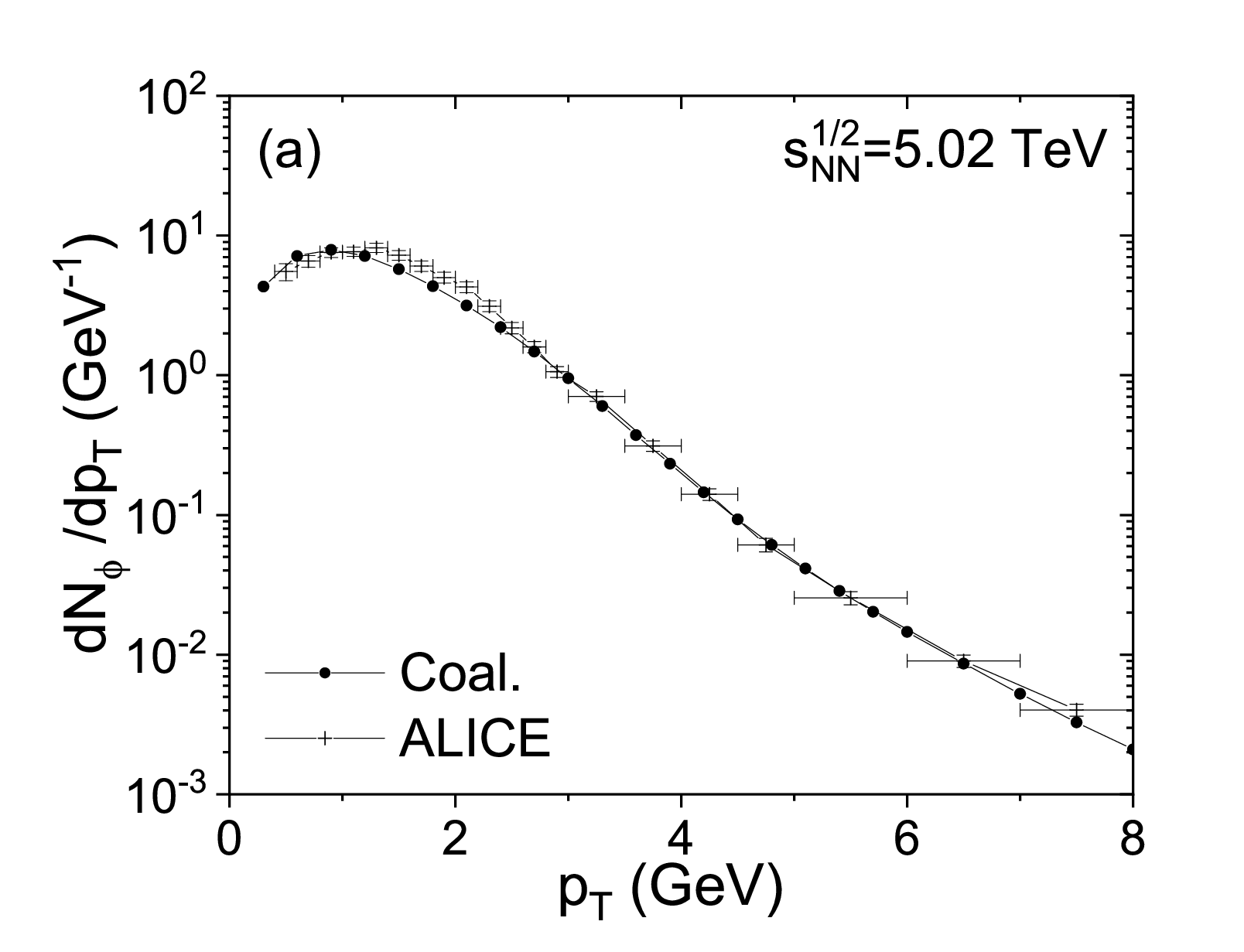}
\includegraphics[width=0.50\textwidth]{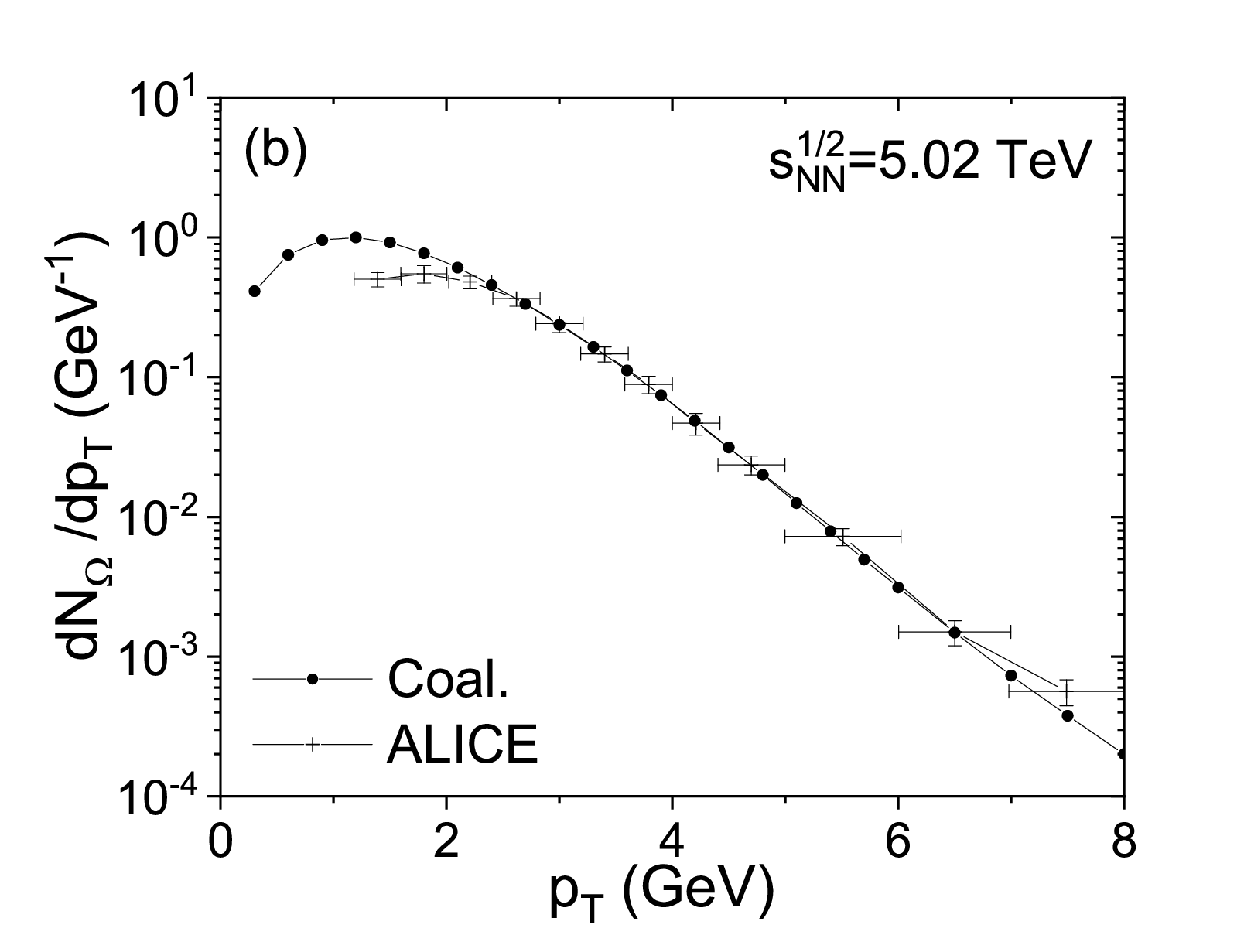}
\end{center}
\caption{Transverse momentum distributions of (a) the $\phi$ meson
and (b) the $\Omega$ baryon, $dN/dp_T$ at mid-rapidity in central
collisions at $\sqrt{s_{NN}}=5.02$ TeV. Also shown are the
transverse momentum distribution of the $\phi$ measured at
$\sqrt{s_{NN}}=5.02$ TeV by ALICE Collaboration
\cite{ALICE:2019xyr}, and that of the $\Omega$ measured at
$\sqrt{s_{NN}}=5.02$ TeV also by ALICE Collaboration
\cite{ALICE:2025cqy}. } \label{pTdistribution_phiOmega}
\end{figure}
Then, using the transverse momentum distribution of strange quarks in Eq. (\ref{d2NsdpT2}), we can calculate the transverse
momentum distribution of the $\phi$ and $\Omega$, Eqs.
(\ref{CoalTransphiW}) and (\ref{CoalTransOmegaW}). Shown in Fig.~\ref{pTdistribution_phiOmega} are the transverse momentum distributions of the $\phi$ meson and the $\Omega$
baryon, $dN/dp_T$ at mid-rapidity in central collisions at
$\sqrt{s_{NN}}=5.02$ TeV. We also show in Fig.~\ref{pTdistribution_phiOmega} the transverse momentum distribution of the $\phi$ measured at $\sqrt{s_{NN}}=5.02$ TeV by ALICE Collaboration \cite{ALICE:2019xyr}, and that of the $\Omega$ measured at $\sqrt{s_{NN}}=5.02$ TeV also by ALICE Collaboration \cite{ALICE:2025cqy}. Here, the set of data for the experimental measurement of the $\phi$ meson transverse momentum distribution \cite{ALICE:2019xyr} has been brought from the hepdata by ALICE
Collaboration, whereas that for the $\Omega$ has been read
directly from the graph in Ref. \cite{ALICE:2025cqy}.

As shown in Fig.~\ref{pTdistribution_phiOmega}(a), the transverse momentum distribution of the $\phi$ meson in the coalescence model, Eq.~\eqref{CoalTransphiW} evaluated with that of strange quarks, Eq.~(\ref{d2NsdpT2}) describes very well the experimental measurements of the transverse momentum distribution of the $\phi$ meson, which is natural because, as mentioned before, the transverse momentum distribution of the strange quarks are
obtained such that it explains the measurement of the $\phi$ meson
transverse momentum distribution at $\sqrt{s_{NN}}=5.02$ TeV at
LHC \cite{Cho:2025lrc}.

On the other hand, it has been found that the evaluation of the
transverse momentum distribution of the $\Omega$ in the
coalescence model, Eq.~\eqref{CoalTransOmegaW} describes the
measurement of that by ALICE Collaboration \cite{ALICE:2025cqy}
reasonably well as shown in Fig.~\ref{pTdistribution_phiOmega}(b),
but overestimates the transverse momentum distribution of the
$\Omega$ in low transverse momentum regions.

The yields of the $\phi$ and the $\Omega$ produced in relativistic
heavy ion collisions at $\sqrt{s_{NN}}=5.02$ TeV energy can be
obtained by carrying out the integration of transverse momentum
distributions shown in Fig.~\ref{pTdistribution_phiOmega} over all
transverse momenta. The yield of the $\phi$ meson is evaluated to
be 13.9 \cite{Cho:2025lrc}, smaller than that of the $\phi$
measured by ALICE Collaboration, 14.937 \cite{ALICE:2019xyr}. On
the other hand, the yield of the $\Omega$ baryon is calculated as
2.09, larger than that measured at mid-rapidity in 0-10$\%$
centralities by ALICE Collaboration, $dN/dy=1.36$
\cite{ALICE:2025cqy}. As mentioned before, the larger yield of the
$\Omega$ baryon compared to the measurement by ALICE Collaboration
is attributable to the overestimation of the transverse momentum
distribution of the $\Omega$ in low transverse momentum regions as
shown in Fig.~\ref{pTdistribution_phiOmega}(b).

Here, the production of the $\phi$ and $\Omega$ purely from the
quark-gluon plasma by strange quark coalescence is taken into
account. Regarding the production of $\phi$ mesons, they are also
produced from the decay of heavier hadrons, especially
charm-strange mesons such as the $D_s$. The $\phi$ is produced
directly from the $D_s$, and indirectly from $D_s^*$,
$D_{s0}^*(2317)$, and $D_{s1}(2460)$ mesons via the $D_s$
\cite{ParticleDataGroup:2024cfk},

\begin{eqnarray}
&& \frac{dN_{\phi}^{tot}}{dp_T}=
\frac{dN_{\phi}}{dp_T}+0.157~\frac{dN_{D_s}^{tot}}{dp_T},
\nonumber \\
&& \frac{dN_{D_s}^{tot}}{dp_T}= \frac{dN_{D_s}}{dp_T}
+\frac{dN_{D_s^*}}{dp_T}+\frac{dN_{D_{s0}^*(2317)}}{dp_T}
\nonumber \\
&& \qquad\quad~+0.820~\frac{dN_{D_{s1}(2460)}}{dp_T}.
\label{feeddowntophi}
\end{eqnarray}
resulting in the total yield of the $\phi$ meson, 14.3
\cite{Cho:2025lrc}, increased by 0.4 from the feed-down
contributions, Eq. (\ref{feeddowntophi}), which is still smaller
than the measurement of the $\phi$ yield by ALICE Collaboration.
Similarly, the production of the $\Omega$ is also possible from
the feed-down contributions such as the $\Omega_c$ and
$\Omega_c(2770)$ \cite{ParticleDataGroup:2024cfk}, expected to
increase the yield of the $\Omega$ further.

\begin{figure}[!t]
\begin{center}
\includegraphics[width=0.50\textwidth]{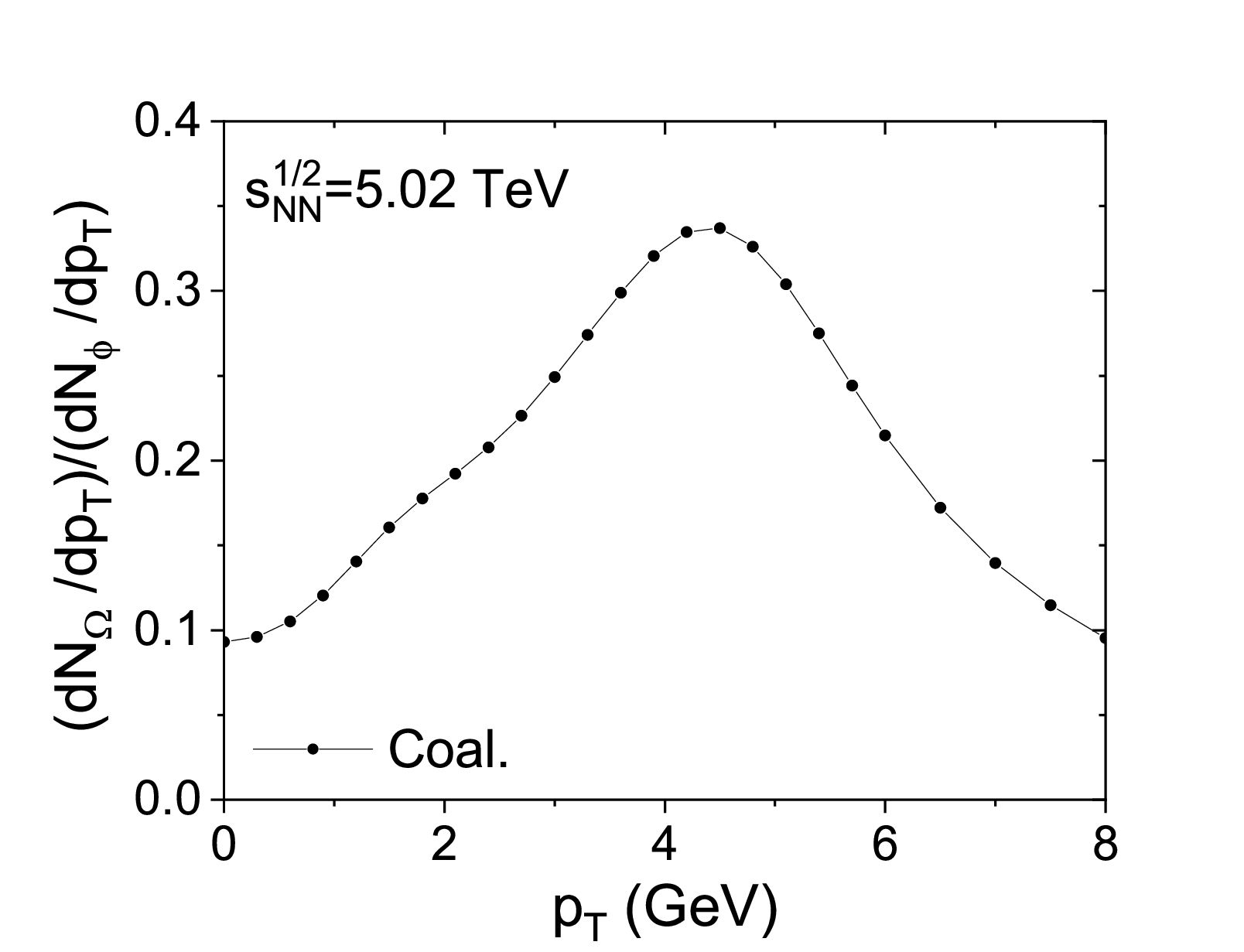}
\end{center}
\caption{Transverse momentum distribution ratio between the
$\Omega$ and the $\phi$, $dN_{\Omega}/dp_T/dN_{\phi}/dp_T$ in
heavy ion collisions at $\sqrt{s_{NN}}=5.02$ TeV. }
\label{pTdistributionratio_phiOmega}
\end{figure}

The comparison of the transverse momentum distribution between the
$\Omega$ and the $\phi$, or the ratio
$dN_{\Omega}/dp_T/dN_{\phi}/dp_T$ is shown in Fig.~\ref{pTdistributionratio_phiOmega}. This ratio is similar to that
between the anti-proton and pion, $\bar{q}\bar{q}\bar{q}/q\bar{q}$
with light quarks $q$, showing the enhanced production of the
anti-proton in the intermediate transverse momentum region
compared to the pion \cite{Greco:2003mm, Greco:2003xt}. For the
ratio between the $\Omega$ and $\phi$, $sss/s\bar{s}$, the peak of
the ratio is also observed in the intermediate transverse momentum
region, but the ratio reaches its maximum value of about 0.35,
implying that the $\Omega$ is always less produced in the entire
range of transverse momentum compared to the $\phi$, though its
production is slightly enhanced in the intermediate transverse
momentum region.

\section{Monte Carlo Hybrid Hadronization for $\phi$ and $\Omega$ Production in Relativistic Heavy-Ion Collisions}
\label{sec3}

\subsection{Parton Distributions}
\label{parton_dist}

Before introducing the hybrid hadronization model, 
we first discuss the parton distributions in central Pb--Pb collisions (0-10 \%) at
$\sqrt{s_{NN}}=5.02$ TeV at LHC.  We classify the partons into two components: thermal partons originating from the QGP and shower partons associated with energetic jets. The thermal component predominantly contributes in the low- and intermediate-$p_T$ regions, whereas the shower component becomes increasingly important at higher transverse momentum.

We employ the blast-wave model to generate the space-time and
momentum-energy distributions of thermal partons in the QGP
produced in central Pb--Pb collisions (0-10\%) at $\sqrt{s_{NN}} =
5.02$ TeV. Assuming boost invariance of the longitudinal momentum
distribution of the thermal partons, we generate them uniformly
within the midrapidity region, $|y|<0.5$. Under the
assumption of the Bjorken correlation $y=\eta$
\cite{Bjorken:1969ja,Bjorken:1982qr}, the longitudinal space-time
coordinates, longitudinal momenta, and energies of thermal partons
are specified by the following relations:
\begin{eqnarray}
&& z=\tau \sinh y, ~ t =\tau \cosh y, \nonumber\\
&& p_z =m_T \sinh\eta, ~ E=m_T \cosh\eta,
\end{eqnarray}
where $m_T$ is the transverse mass of the quark or antiquark, defined as $m_T =\sqrt{p_T ^2
+ m_{q,\bar{q}} ^2}$, and $\tau$ is the proper time. Due to the
collective expansion of the QGP, the thermal partons are assumed to be boosted
by a radial flow velocity $\vec v_{T}=\beta_0
\left(\vec r_{T}/R_{\perp}\right)$, where $R_{\perp}$ is the
transverse radius of the QGP fireball at hadronization,
$\vec r_T$ denotes the transverse positions of
partons, and $\beta_0$ is the collective flow velocity of the QGP
medium. The transverse momentum distribution of thermal partons
is given \cite{Greco:2003mm} in the following
equation:
\begin{eqnarray}
\frac{dN_{q,\bar{q}}}{d^2\vec r_{T}d^{2}\vec p_{T}}=
\frac{g_{q,\bar{q}}\tau
m_{T}}{(2\pi)^3}\exp\Big(-\frac{\gamma_{T}(m_{T}-
\vec p_{T}\cdot\vec v_{T})\mp\mu_{b}}{T}\Big),\nonumber \\
\label{thermq}
\end{eqnarray}
where spin-color degeneracies of partons $g_{q,\bar{q}}$ are 6 for
quark and anti-quark and 16 for gluons. We use
$m_{q,\bar{q}}=m_{g} = 340$ MeV for the masses of light quarks and
gluons and $m_{s,\bar{s}}=486$ MeV. Adopting the fireball
parameters from Ref. \cite{Minissale:2023dct}, we take $R_{\perp}
= 14.1$ fm  for a transverse radius of the fireball at proper time
$\tau = 8$ fm/c, $V\approx 5000 ~\rm{fm}^3$ for
its volume, $\mu_b=0$ for the baryon chemical potential, and
$\beta_0 =0.7 c$ for the collective flow velocity. The QGP temperature is assumed to be $T=175$ MeV. Although the transverse geometry of the QGP is generally anisotropic, as reflected by the finite elliptic flow observed experimentally, we neglect this anisotropy because the present analysis considers observables integrated over the full azimuthal angle, $2\pi$.

\begin{figure}[!t]
\begin{center}
\includegraphics[width=0.50\textwidth]{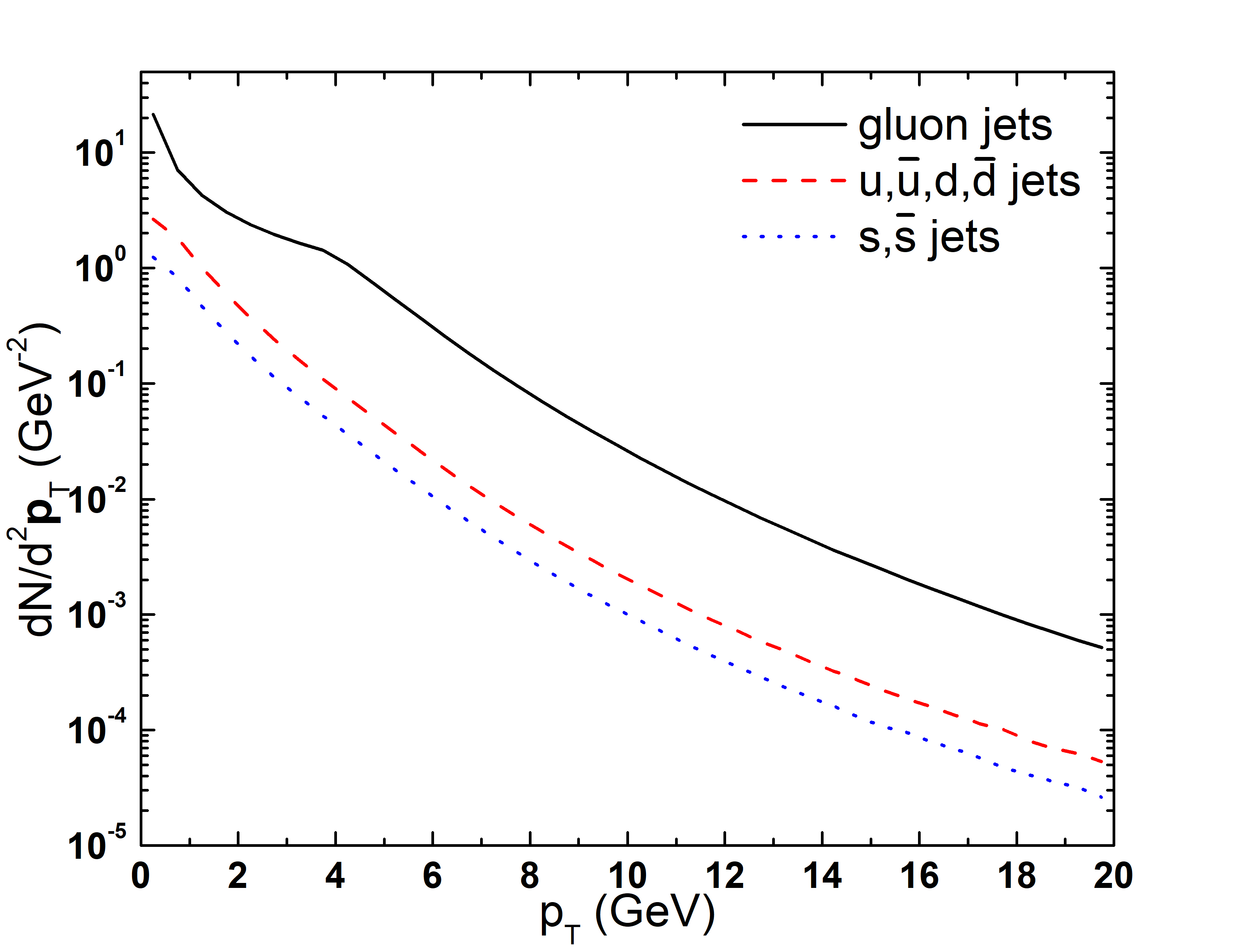}
\end{center}
\caption{Transverse momentum distribution of gluon (sold line),
light quark (dashed line), and strange quark (dotted line) jets in
the central Pb--Pb collisions (0-10\%) at $\sqrt{s_{NN}}=5.02$
TeV.} \label{pTdistribution_jetpartons}
\end{figure}

Before generating parton showers, we use HIJING (Heavy Ion Jet
INteraction Generator)~\cite{Wang:1991hta,Gyulassy:1994ew} to
generate hard-scattered partons in central (0--10\%) Pb--Pb
collisions at $\sqrt{s_{NN}}=5.02$ TeV using the standard HIJING
1.4 settings, except that jet quenching is switched off. HIJING
provides a perturbative QCD-based description of hard parton
production embedded in a soft string excitation framework,
including nuclear shadowing and multiple scattering effects
relevant to heavy-ion collisions. The resulting hard partons
therefore serve as unquenched initial jet partons, which are
subsequently evolved into medium-modified parton showers using
Q-PYTHIA. Fig.~\ref{pTdistribution_jetpartons} shows the
transverse-momentum distributions of the initial jet partons
generated by HIJING.
The medium modification of the parton shower is simulated using
Q-PYTHIA \cite{Armesto:2009fj}, a Monte Carlo event generator that
extends PYTHIA \cite{Sjostrand:1984ic} by incorporating
medium-induced gluon radiation based on the BDMPS--Z formalism
\cite{Baier:1996kr,Zakharov:1996fv}. Each jet parton generated by
HIJING is used as the initial parton for Q-PYTHIA, which evolves
the parton shower through the dense QCD medium by enhancing the
parton splitting probabilities according to the local transport
coefficient, $\hat{q}$. This procedure produces medium-modified
parton showers whose momentum and flavor distributions naturally
incorporate the effects of jet quenching in relativistic heavy-ion
collisions. In the present study, we adopt
$\hat{q}=2~\mathrm{GeV}^2/\mathrm{fm}$ and a medium path length of
$L=5~\mathrm{fm}$. The resulting shower partons exhibit the
expected medium-induced transverse momentum broadening and energy
softening, providing a realistic description of jet-medium
interactions that is essential for modeling the production of
strange and multi-strange hadrons at intermediate transverse
momentum.
\begin{figure}[!t]
\begin{center}
\includegraphics[width=0.40\textwidth]{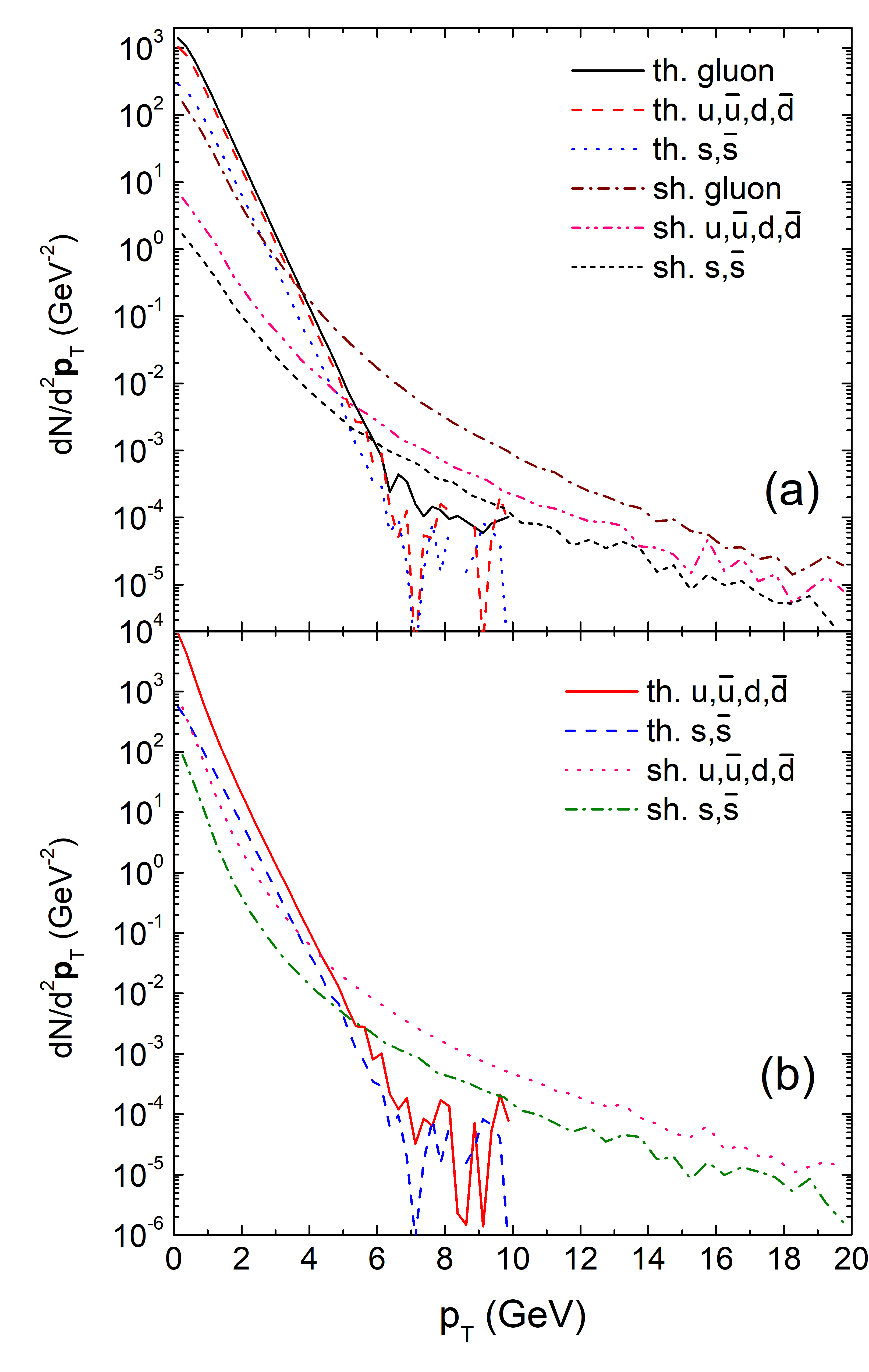}
\end{center}
\caption{Transverse-momentum distributions of thermal (solid,
long-dashed, and dotted lines) and shower (dot-dashed,
dash-double-dotted, and short-dashed lines) gluons, light quarks,
and strange quarks produced in central (0-10 \%) Pb--Pb collisions
at $\sqrt{s_{NN}}= 5.02$ TeV before gluon decay (panel (a)). After
gluon decay, the corresponding thermal (solid and long-dashed
lines) and shower (dotted and dot-dashed lines) distributions of
light and strange quarks are shown in panel (b).}
\label{pTdistribution_thsh}
\end{figure}
Fig.~\ref{pTdistribution_thsh} (a) shows the
transverse-momentum distributions of thermal partons generated
using the blast-wave model and shower partons generated with
Q-PYTHIA, following the procedure described above. Initial hard
partons generated by HIJING are assumed to be uniformly
distributed in the transverse plane ($z=0$) at the initial time
$t=0$. During the subsequent parton shower evolution, each
propagating parton moves with velocity $\vec v=\vec p/E$ over a
lifetime $\tau = E/Q^2$, where $Q$ is its virtuality. After each
branching, the daughter partons are assigned their own space--time
coordinates according to their propagation histories. This
procedure enables us to reconstruct the full space--time
information of all shower partons, which serves as the input to
the event-by-event Monte Carlo recombination model.

Since hadrons in the recombination model are formed from quarks and antiquarks, gluons are converted into quark--antiquark pairs by assigning them random virtualities in the range $0.68$--$1.3$ GeV. If the assigned virtuality is below the threshold $2m_{s,\bar{s}}=0.97$ GeV, the gluon is converted into a light quark--antiquark pair. For virtualities above this threshold, the gluon is converted into a strange quark--antiquark pair. The details of the gluon conversion mechanism are described in Ref.~\cite{Han:2016uhh}.
Fig.~\ref{pTdistribution_thsh} (b) 

presents the transverse-momentum
distribution of thermal and shower quarks obtained after
implementing the gluon decay procedure described above. The
daughter quark and antiquark produced from gluon conversion are
assigned space--time coordinates using the same propagation
procedure adopted for the parton shower evolution. The parent
gluon first propagates according to its momentum and virtuality,
and the quark--antiquark pair is generated at the corresponding
decay point. The daughter partons then continue to propagate with
their respective momenta and virtualities until hadronization.

\subsection{Monte Carlo Hadronization Framework}
\label{sec:MonteCarloFramework}

The thermal and medium-modified shower partons generated in
Sec.~\ref{parton_dist} are hadronized event by event within a
Monte Carlo hybrid hadronization framework that incorporates
quark recombination with Lund string
fragmentation. The framework provides a unified description of
hadron production over a broad transverse-momentum range by
incorporating recombination among thermal and shower partons
together with the fragmentation of remnant shower partons. In this
picture, recombination is expected to dominate hadron production
at low and intermediate transverse momentum, whereas string
fragmentation becomes increasingly important at high transverse
momentum.

Candidate hadrons are constructed event by event from possible
combinations of constituent quarks with the appropriate flavor
content. Mesons are formed from quark--antiquark pairs, whereas
baryons are formed from three quarks. Depending on the origin of the constituent partons, meson recombination proceeds through thermal--thermal (th-th), thermal--shower (th-sh),
and shower--shower (sh-sh) channels. For baryons, the
corresponding recombination channels involve different
combinations of three thermal and shower quarks:
thermal--thermal--thermal (th-th-th), thermal--thermal--shower
(th-th-sh), thermal--shower--shower (th-sh-sh), and
shower--shower--shower (sh-sh-sh). The relative contributions of
these channels are determined dynamically by the event-by-event
phase-space distributions of the thermal and shower partons.

The momentum distributions of mesons and baryons produced through
quark recombination are generally expressed as
\cite{Greco:2003xt,Han:2016uhh}
\begin{eqnarray}
\frac{dN_M}{d^3{\vec P}_M} &=& g_M \int d^3{\vec x}_{q}d^3{\vec
p}_{q} d^3{\vec x}_{\bar q}d^3{\vec p}_{\bar q} f_q({\vec
x}_q,{\vec p}_q) f_{\bar q}({\vec x}_{\bar q},{\vec p}_{\bar q})
\nonumber\\
&&\times W_M({\vec r},{\vec k}) \delta^{(3)} ({\vec P}_M-{\vec
p}_q-{\vec p}_{\bar q}), \label{coalMprob}
\end{eqnarray}
and
\begin{eqnarray}
\frac{dN_B}{d^3{\vec P}_B} &=& g_B \int d^3{\vec x}_{1}d^3{\vec
p}_{1} d^3{\vec x}_{2}d^3{\vec p}_{2} d^3{\vec x}_{3}d^3{\vec
p}_{3} f_{q_1}({\vec x}_1,{\vec p}_1)
\nonumber\\
&&\times f_{q_2}({\vec x}_2,{\vec p}_2) f_{q_3}({\vec x}_3,{\vec
p}_3) W_B({\vec r}_1,{\vec k}_1;{\vec r}_2,{\vec k}_2)
\nonumber\\
&&\times \delta^{(3)} ({\vec P}_B-{\vec p}_1-{\vec p}_2-{\vec
p}_3), \label{coalBprob}
\end{eqnarray}
where $g_M$ and $g_B$ denote the statistical color--spin
degeneracy factors for mesons and baryons, respectively, with
their values specified in Sec.~\ref{sec2}. The relative
coordinates and momenta of the constituent partons, ${\vec r}$,
${\vec k}$, ${\vec r}_1$, ${\vec r}_2$, ${\vec k}_1$, and ${\vec
k}_2$, are defined in Eqs.~(\ref{relcoordmeson}) and
(\ref{relcoordbaryon}). The Wigner functions $W_M$ and $W_B$,
given in Eq.~(\ref{WignerS}), characterize the phase-space overlap
of the constituent partons and determine the corresponding
recombination probabilities. In the present implementation,
explicit color flow is not tracked during the recombination
procedure. Instead, the probability for the constituent partons to
form color-singlet hadrons is incorporated statistically through
the factors $g_M$ and $g_B$. We use the same width parameters, $\sigma_{\phi}=0.38$~fm, $\sigma_{\Omega_1}=3.04$~fm, and $\sigma_{\Omega_2}=2.63$~fm, as in Sec.~\ref{sec2}.

For each possible hadron candidate, the corresponding Wigner
function is evaluated using the relative coordinates and momenta
of its constituent partons. A random number uniformly distributed
between 0 and 1 is then generated and compared with the value of
the Wigner function. If the Wigner-function value exceeds the
generated random number, recombination is accepted and the
constituent partons are removed from the available parton
ensemble. Otherwise, the candidate is rejected and the partons
remain available for subsequent recombination attempts. This
acceptance--rejection procedure is repeated until no further
recombination channels are available.

After the recombination stage, the remaining shower partons are
prepared for string fragmentation. Prior to recombination, gluons
are allowed to undergo nonperturbative splitting into
quark--antiquark pairs so that the resulting quarks and antiquarks
can participate in the recombination process. If neither daughter
parton from such a gluon splitting participates in recombination,
the splitting is reversed and the original gluon is restored. The
remaining shower quarks, antiquarks, restored gluons, and diquarks
are then reconnected into color-singlet strings of the forms
$(q,\bar q)$, $(q,g,g,\ldots,\bar q)$, $(q,qq)$, and
$(q,g,g,\ldots,qq)$, where $qq$ denotes a diquark. These strings
are subsequently hadronized through the Lund string fragmentation
model implemented in PYTHIA~6.3 \cite{Han:2016uhh,Sjostrand:2006za}. In
contrast, thermal partons that do not participate in recombination
are not passed to the fragmentation stage. The resulting framework
thus provides an event-by-event description of hadron production
through the combined effects of quark recombination and remnant
shower fragmentation.

\subsection{Transverse-Momentum Spectra of the $\phi$ Meson and
$\Omega$ Baryon} \label{sec:ptspectra}

\begin{figure}[!h]
\begin{center}
\includegraphics[width=0.50\textwidth]{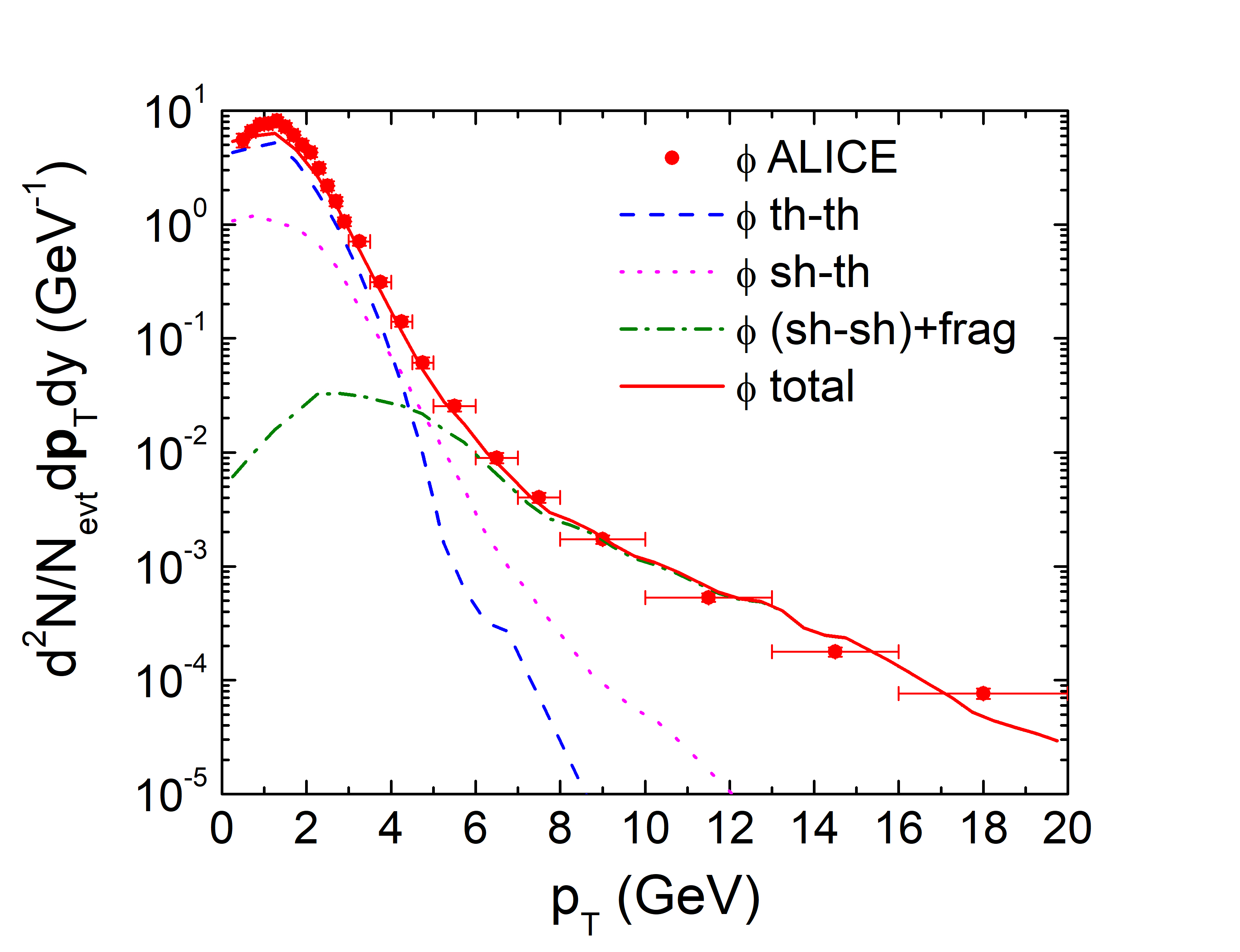}
\includegraphics[width=0.50\textwidth]{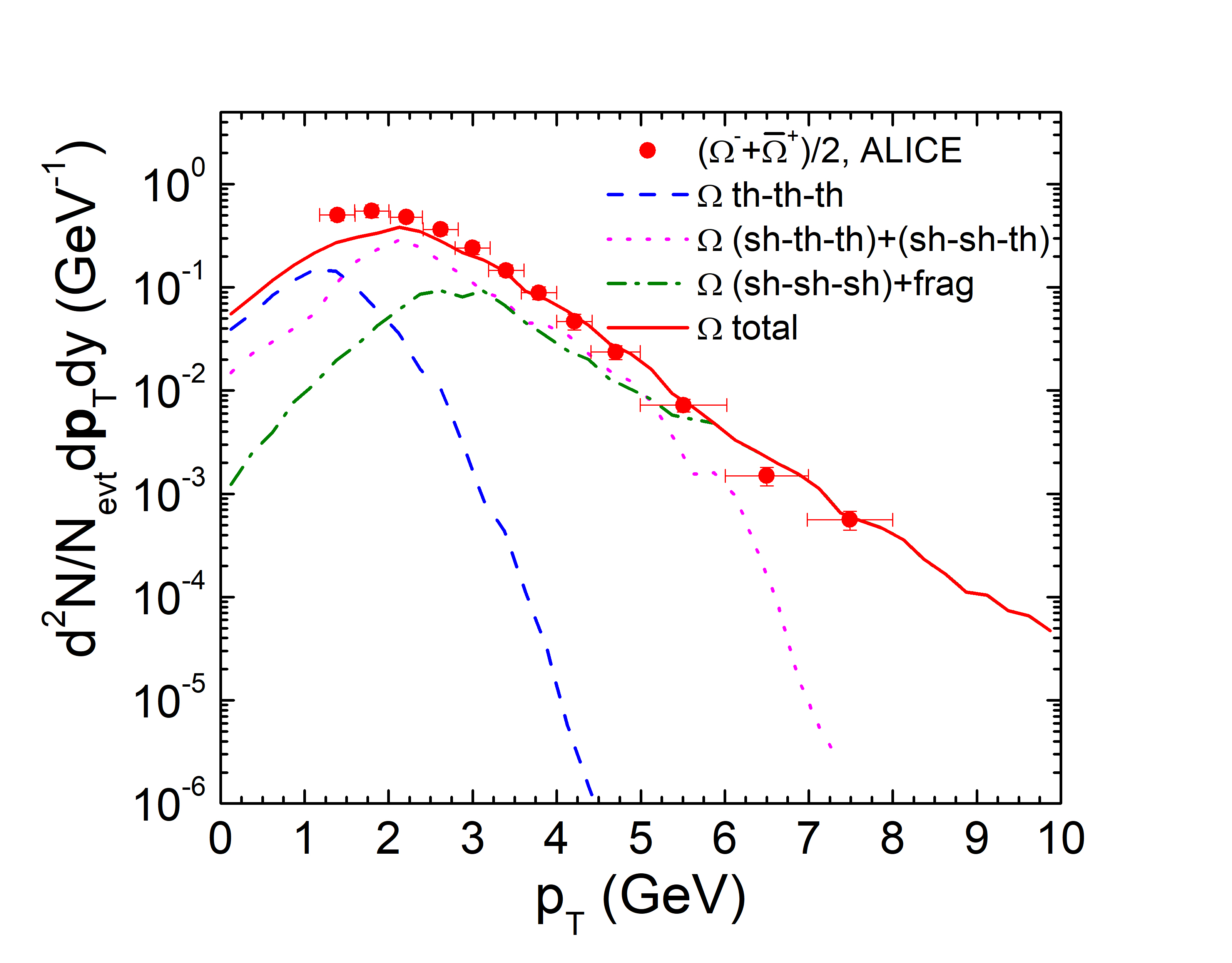}
\end{center}
\caption{Transverse-momentum spectra of the $\phi$ meson (upper
panel) and the $\Omega$ baryon (lower panel) at midrapidity
($|y|<0.5$) in central (0--10\%) Pb--Pb collisions at
$\sqrt{s_{NN}}=5.02$ TeV. For the $\phi$ meson, the dashed and
dotted curves represent the (th-th) and (th-sh)
recombination contributions, respectively, while the dash-dotted
curve represents the combined (sh-sh) recombination and remnant
shower fragmentation contributions. For the $\Omega$ baryon, the
dashed curve represents the (th-th-th) contribution, the dotted
curve represents the combined (th-th-sh) and (th-sh-sh)
contributions, and the dash-dotted curve represents the combined
(sh-sh-sh) recombination and remnant shower fragmentation
contributions. The solid curves represent the total calculated
spectra. The filled circles represent the corresponding
measurements by the ALICE Collaboration
\cite{ALICE:2019xyr,ALICE:2025cqy}.} \label{phi_and_Omega}
\end{figure}

Fig.~\ref{phi_and_Omega} shows the calculated
transverse-momentum spectra of the $\phi$ meson (upper panel) and
the $\Omega$ baryon (lower panel) at midrapidity ($|y|<0.5$) in
central (0--10\%) Pb--Pb collisions at $\sqrt{s_{NN}}=5.02$ TeV.
For the $\phi$ meson, the thermal--thermal (th-th)
and thermal--shower (th-sh) recombination contributions are shown
separately, while the shower--shower (sh-sh) contribution is
combined with the fragmentation of unrecombined remnant shower
partons. For the $\Omega$ baryon, the thermal--thermal--thermal
(th-th-th) contribution is shown separately, while the mixed
thermal--shower channels, (th-th-sh) and (th-sh-sh), are combined.
The shower--shower--shower (sh-sh-sh) contribution is likewise
shown together with the fragmentation of remnant shower partons.
These groupings are adopted for clarity in presenting the relative
contributions of the different hadronization mechanisms. The solid
curves represent the total spectra obtained by summing all
contributions, while the filled circles represent the
corresponding measurements by the ALICE Collaboration
\cite{ALICE:2019xyr, ALICE:2025cqy}.

The present hybrid hadronization model reproduces the overall
features of both the $\phi$-meson and $\Omega$-baryon spectra over
a broad transverse-momentum range. At low transverse momentum,
hadron production is governed primarily by recombination among
thermal strange quarks, corresponding to the
(th-th) channel for the $\phi$ meson and the
(th-th-th) channel for the $\Omega$ baryon. This behavior reflects
hadronization from the collectively expanding thermal medium. As
the transverse momentum increases, recombination involving both
thermal and shower partons becomes progressively more important.
The (th-sh) channel for the $\phi$ meson and the mixed (th-th-sh)
and (th-sh-sh) channels for the $\Omega$ baryon provide an
important contribution in the intermediate-$p_T$ region,
illustrating the role of recombination between energetic shower
partons and strange quarks from the surrounding thermal medium. At
still higher transverse momentum, shower-dominated recombination
and remnant shower fragmentation become increasingly important.

The calculated spectra underestimate the measured yields in the
low-$p_T$ region, with a more pronounced deviation for the
$\Omega$ baryon. Since hadron production in this region is
dominated by thermal recombination, the discrepancy may reflect
limitations of the simplified description of the thermal
strange-quark phase-space distribution employed in the present
blast-wave parameterization. In particular, $\Omega$ production
requires the simultaneous recombination of three strange quarks
and is therefore more sensitive to the thermal strange-quark
phase-space density than $\phi$-meson production, which requires
an $s\bar{s}$ pair. Nevertheless, the overall behavior of both
measured spectra is reasonably reproduced within the same
event-by-event hybrid hadronization framework.

\begin{figure}[h]
\begin{center}
\includegraphics[width=0.5\textwidth]{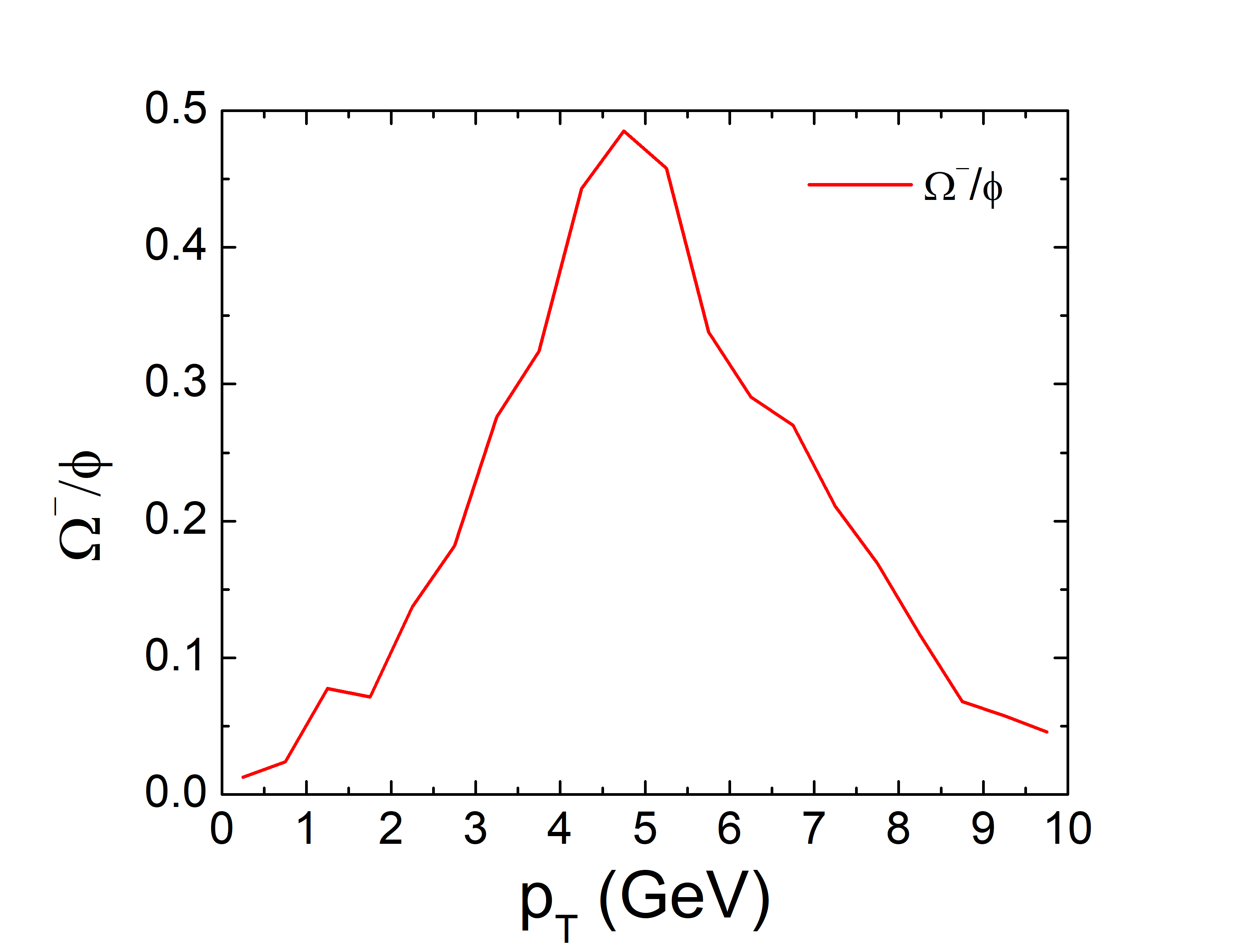}
\end{center}
\caption{ Predicted transverse-momentum dependence of the
$\Omega/\phi$ ratio at midrapidity ($|y|<0.5$) in
central (0--10\%) Pb--Pb collisions at $\sqrt{s_{NN}}=5.02$ TeV.
The ratio is calculated from the total $\Omega$-baryon and
$\phi$-meson yields, including contributions from quark
recombination and remnant shower fragmentation.}
\label{phitoOmegaMC}
\end{figure}

Fig.~\ref{phitoOmegaMC} shows the predicted transverse-momentum
dependence of the $\Omega^{-}/\phi$ ratio obtained from the total
calculated hadron yields.

The $\Omega/\phi$ ratio increases with transverse
momentum, reaches a maximum value of approximately 0.48 at
$p_T\approx5$ GeV, and then gradually decreases toward higher
transverse momentum. The enhancement in the intermediate-$p_T$
region reflects the characteristic baryon-to-meson enhancement
associated with quark recombination in a dense partonic medium. In
particular, the different momentum dependence of meson and baryon
recombination leads to an enhanced relative production of the
$\Omega$ baryon in this region. At higher transverse momentum,
shower-dominated hadronization and remnant shower fragmentation
become increasingly important, resulting in a decrease of the
$\Omega/\phi$ ratio. Since no experimental
measurement of this ratio is currently available for Pb--Pb
collisions at $\sqrt{s_{NN}}=5.02$ TeV, the present result
provides a prediction that can be tested in future measurements at
the LHC.

\section{Conclusions}
\label{sec4}

We have studied in this works the production of the $\phi$ and
$\Omega$ in heavy ion collisions at $\sqrt{s_{NN}} = 5.02$ TeV by
employing two complementary approaches. Starting from the
discussion about the $\phi$ and $\Omega$ production based on a
coalescence model, we have investigated in more detail the
production of the $\phi$ and $\Omega$ within a hybrid
hadronization framework that combines quark recombination with
Lund string fragmentation. We have calculated the transverse
momentum distribution of the $\phi$ and $\Omega$, and have also
evaluated the ratio of their transverse momentum distributions.

In the coalescence model formalism, we have adopted the
predetermined strange-quark phase-space distribution for strange
quarks in heavy ion collisions at $\sqrt{s_{NN}} = 5.02$ TeV, and
have calculated transverse-momentum distribution and yield of the
$\Omega$. We show that our calculations of the $\Omega$
transverse-momentum distribution agree reasonably well with that
measured by ALICE Collaboration, while somewhat overestimating the $\Omega$ transverse-momentum distribution in the
low transverse-momentum region. 
The transverse momentum
distribution ratio between the $\Omega$ and $\phi$ has been found
to be enhanced in the intermediate transverse momentum regions,
similar to that between the antiproton and pion.

In a hybrid hadronization framework, we have extended the hybrid
hadronization framework for the hadronization of perturbative
parton showers in vacuum to take into account the production of
$\phi$ and $\Omega$ in a QGP medium. In this extension, thermal
partons from the QGP have been combined with medium-modified
shower partons originating from energetic jets through quark
recombination, while shower partons that remain unrecombined have
been subsequently hadronized through the Lund string fragmentation
model. This event-by-event hybrid framework has been shown to
provide a unified description of hadron production over a broad
transverse-momentum range by incorporating both recombination and
fragmentation.

We find that this unified treatment naturally connects the
recombination-dominated intermediate-$p_T$ region with the
fragmentation-dominated high-$p_T$ region, providing a continuous
description of hadron production over a broad transverse-momentum
range. Moreover, we show that we can decompose in a hybrid
hadronization framework the recombination contributions to the
thermal--thermal (th-th), thermal--shower (th-sh) and the
shower--shower (sh-sh) for the production of the $\phi$, and the
thermal--thermal--thermal (th-th-th), thermal--shower channels,
(th-th-sh) and (th-sh-sh), and shower--shower--shower (sh-sh-sh)
for the production of the $\Omega$, respectively, enabling to
investigate in detail the relative contributions of the different
hadronization mechanisms.

The resulting transverse-momentum spectra of $\phi$ and $\Omega$
in this approach demonstrate that thermal recombination dominates
strange-hadron production at low and intermediate transverse
momenta, whereas fragmentation of remnant shower partons becomes
increasingly important at high $p_T$. The combined contribution
from recombination and fragmentation reproduces the overall shape
of the measured $\phi$ and $\Omega$ spectra significantly better
than either mechanism alone, indicating that both hadronization
mechanisms are essential for describing strange-hadron production
in relativistic heavy-ion collisions.

We expect that the present framework provides a flexible
foundation for future investigations. Since the model preserves
the complete event-by-event space-time information of all partons,
it can be straightforwardly extended to calculate anisotropic flow
coefficients, nuclear modification factors, baryon-to-meson
ratios, and multi-particle correlations for various hadrons.
Furthermore, the same framework can be applied to investigate the
production of heavy-flavor hadrons and exotic multiquark states,
thereby providing a unified description of hadronization in a
strongly interacting QCD matter.

\section*{Acknowledgments}
K.C.H. acknowledges support from the 2025 Faculty Enhancement
Program at Prairie View A \& M University. This work was supported
by the National Research Foundation of Korea (NRF) grant funded by
the Korea government (MSIT) No. RS-2023-00280831 (S.C.),  No. RS-2023-NR077232 (S.H.L.).

%\bibliographystyle{Science}
%\bibliographystyle{h-physrev}

% \bibliography{references}

\end{document}